\documentclass[final,5p,times,twocolumn]{elsarticle}
\usepackage{amssymb}
\usepackage[subrefformat=parens]{subcaption}
\usepackage{amsmath}
\usepackage{stfloats}

\begin{document}
\begin{frontmatter} 



\title{Hyperon--nucleon interaction through the $K^-d\to\pi\Lambda N$ reaction}


\author{Shunsuke Yasunaga} 
\author{Daisuke Jido}
\affiliation{organization={Department of Physics, Institute of Science Tokyo},
            addressline={2-12-1 Ookayama, Meguro-ku}, 
            city={Tokyo},
            postcode={152-8551}, 
            country={Japan}}

\begin{abstract}
The hyperon--nucleon interaction is investigated through the final-state interaction in the $K^-d\to\pi^-\Lambda p$ reaction.
We focus on the $\Lambda N$--$\Sigma N$ coupled-channel interaction, which produces characteristic structures around the $\Sigma N$ thresholds in the $\Lambda p$ invariant mass spectrum.
The spin-triplet $\Sigma N\to\Lambda p$ conversion amplitude is constructed within the $K$-matrix formalism using scattering lengths in the isospin basis.
We first examine the dependence of the conversion amplitude on the $\Sigma N$ scattering lengths and find that the threshold structure is particularly sensitive to the sign of the real part of the $I=1/2$ scattering length.
We then calculate the $\Lambda p$ invariant mass spectrum of the $K^-d\to\pi^-\Lambda p$ reaction, including the contributions from the background diagrams.
The resulting spectra show characteristic structures around the $\Sigma N$ thresholds, whose shapes depend on the choice of the interaction parameters. These results suggest that the $\Lambda p$ invariant mass spectrum can serve as a useful observable for constraining the $\Lambda N$--$\Sigma N$ coupled-channel interaction.
\end{abstract}




\end{frontmatter}



\section{Introduction}\label{sec1}
The hyperon--nucleon ($YN$) interaction plays a fundamental role in our understanding of strangeness nuclear physics. In particular, it is expected to significantly affect the properties of dense baryonic matter such as neutron stars~\cite{chatterjee_2016} However, in contrast to the nucleon--nucleon interaction, experimental constraints on the $YN$ interaction remain limited due to the difficulty of direct scattering experiments. Alternative approaches, including hypernuclear spectroscopy~\cite{akaishi_2002}, femtoscopy in nuclear collisions~\cite{acharya_2019,acharya_2020}, and few-body reactions~\cite{budzanowski_2010} are essential for extracting information on the $YN$ interaction.

The $K^-d\to\pi\Lambda N$ reaction provides a unique opportunity to investigate the $YN$ interaction through the final-state interaction. In previous works~\cite{iizawa_2022,yasunaga_2025}, isospin symmetry breaking in the $\Lambda N$ interaction was studied by analyzing the ratio of the invariant mass spectra of the $K^-d\to\pi^-\Lambda p$ and $K^-d\to\pi^0\Lambda n$. In this work, as a continuation of our previous proceedings work~\cite{yasunaga_2025a}, we focus on the $\Lambda N$--$\Sigma N$ coupled-channel interaction. A prominent feature of this system is the cusp structure at the $\Sigma N$ threshold in the $\Lambda N$ invariant mass. This originates from the opening of the coupled channel and is sensitive to the off-diagonal transition amplitude between $\Lambda N$ and $\Sigma N$ states. Recently, the J-PARC E90 experiment~\cite{ichikawa_2022} has been ongoing to investigate the $YN$ interaction through the $K^-d\to\pi^-\Lambda p$ reaction. With a view toward comparison with such high-resolution experiments, we calculate the $\Lambda p$ invariant mass spectrum of the $K^-d\to\pi^-\Lambda p$ reaction, focusing on the $\Lambda N$--$\Sigma N$ coupled-channel interaction.

\begin{figure*}[!t]
\begin{subfigure}{0.48\textwidth}
\includegraphics[width=\textwidth]{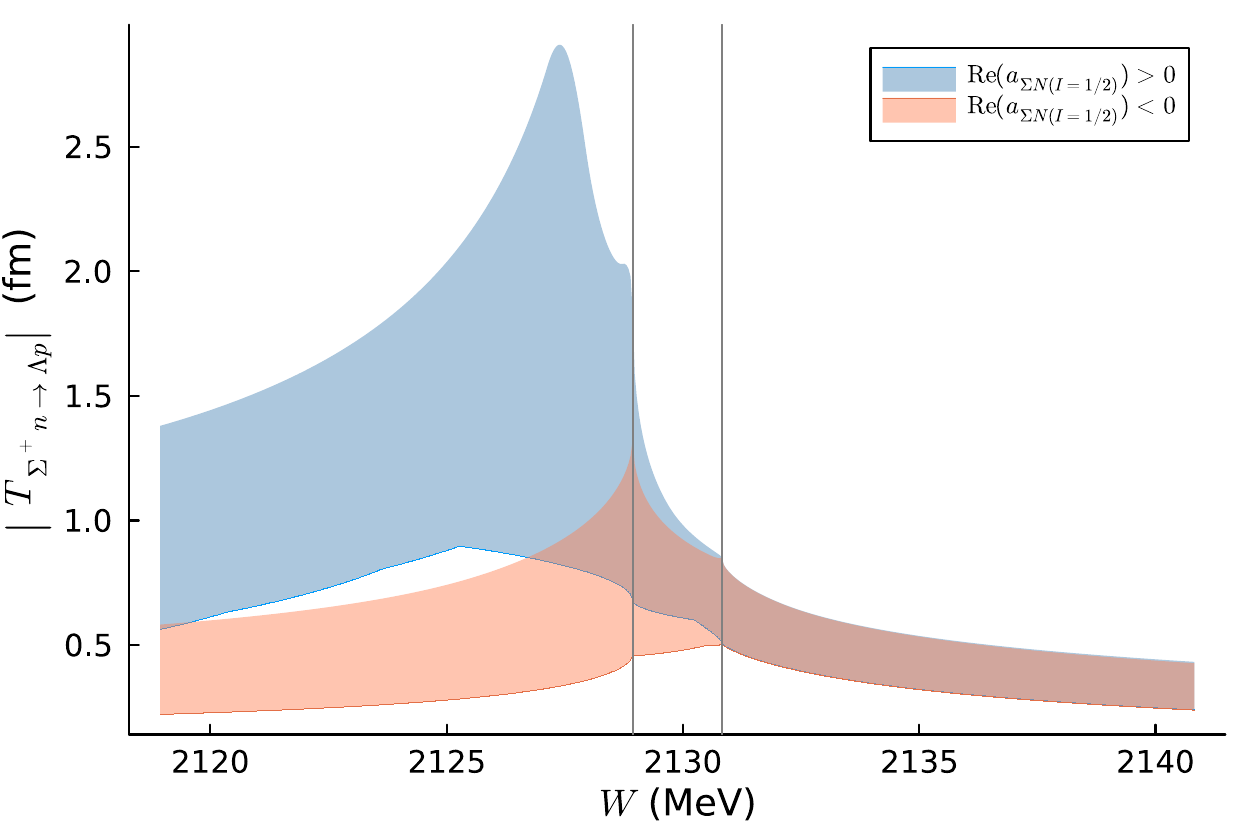}
\subcaption{Envelope obtained by varying $\mathrm{Re}(a_{\Sigma N(I=1/2)}) \in [-4.0,-1.0]$~fm and $[1.0,4.0]$~fm, $\mathrm{Im}(a_{\Sigma N(I=1/2)}) \in [1.0,4.0]$~fm, and $a_{\Sigma N(I=3/2)} \in [-1.0,1.0]$~fm simultaneously. The blue (red) band corresponds to $\mathrm{Re}(a_{\Sigma N(I=1/2)}) > 0$ ($< 0$).}
\end{subfigure}
\hfill
\begin{subfigure}{0.48\textwidth}
\includegraphics[width=\textwidth]{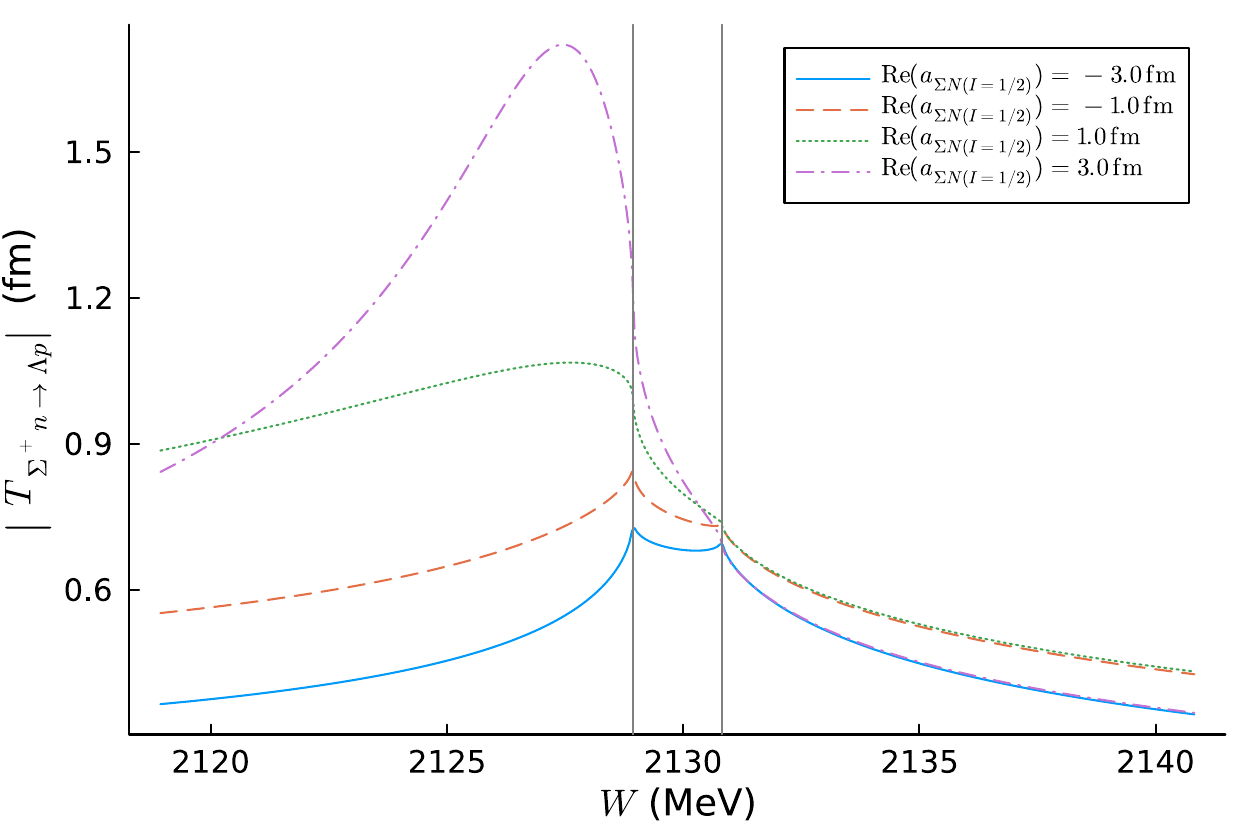}
\subcaption{Dependence on the real part of the $I=1/2$ scattering length, with $\mathrm{Im}(a_{\Sigma N(I=1/2)}) = 2.0$~fm fixed.}
\end{subfigure}
\hfill
\begin{subfigure}{0.48\textwidth}
\includegraphics[width=\textwidth]{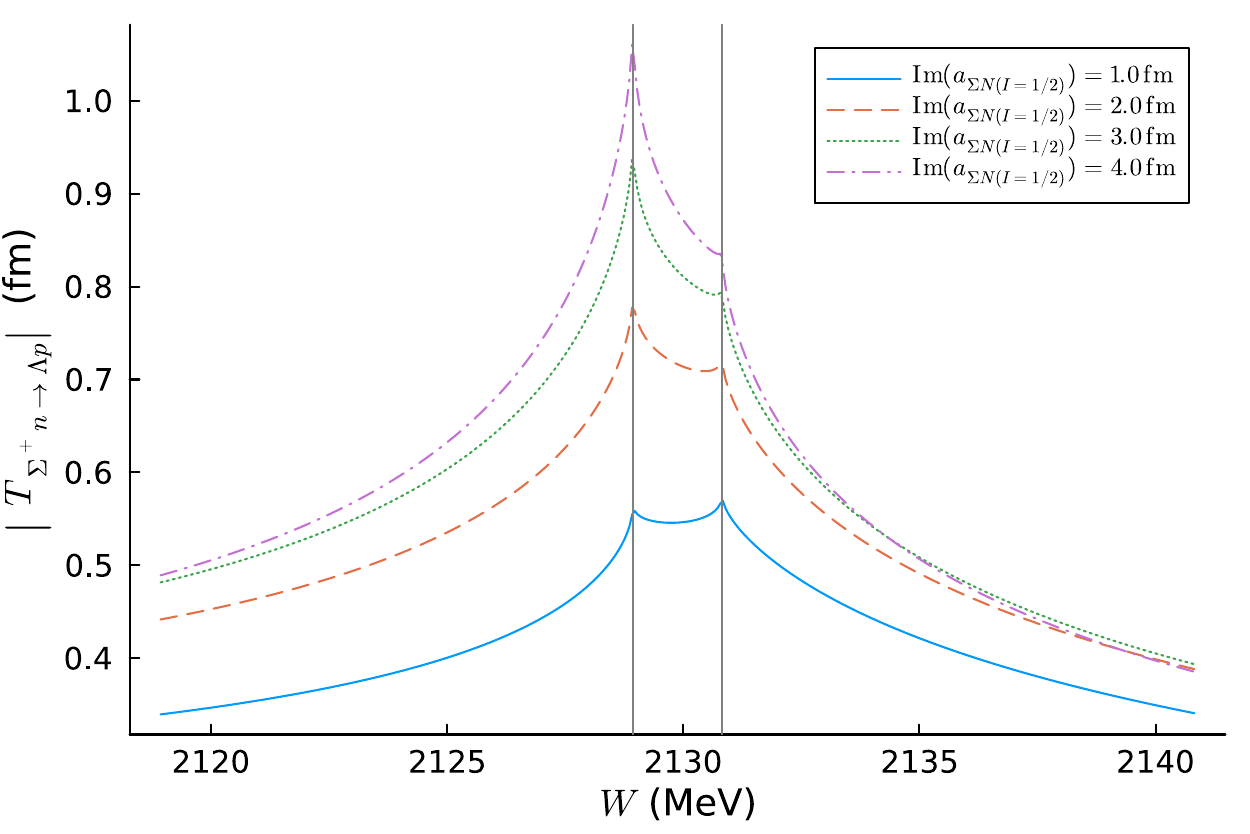}
\subcaption{Dependence on the imaginary part of the $I=1/2$ scattering length, with $\mathrm{Re}(a_{\Sigma N(I=1/2)}) = -2.0$~fm fixed.}
\end{subfigure}
\hfill
\begin{subfigure}{0.48\textwidth}
\includegraphics[width=\textwidth]{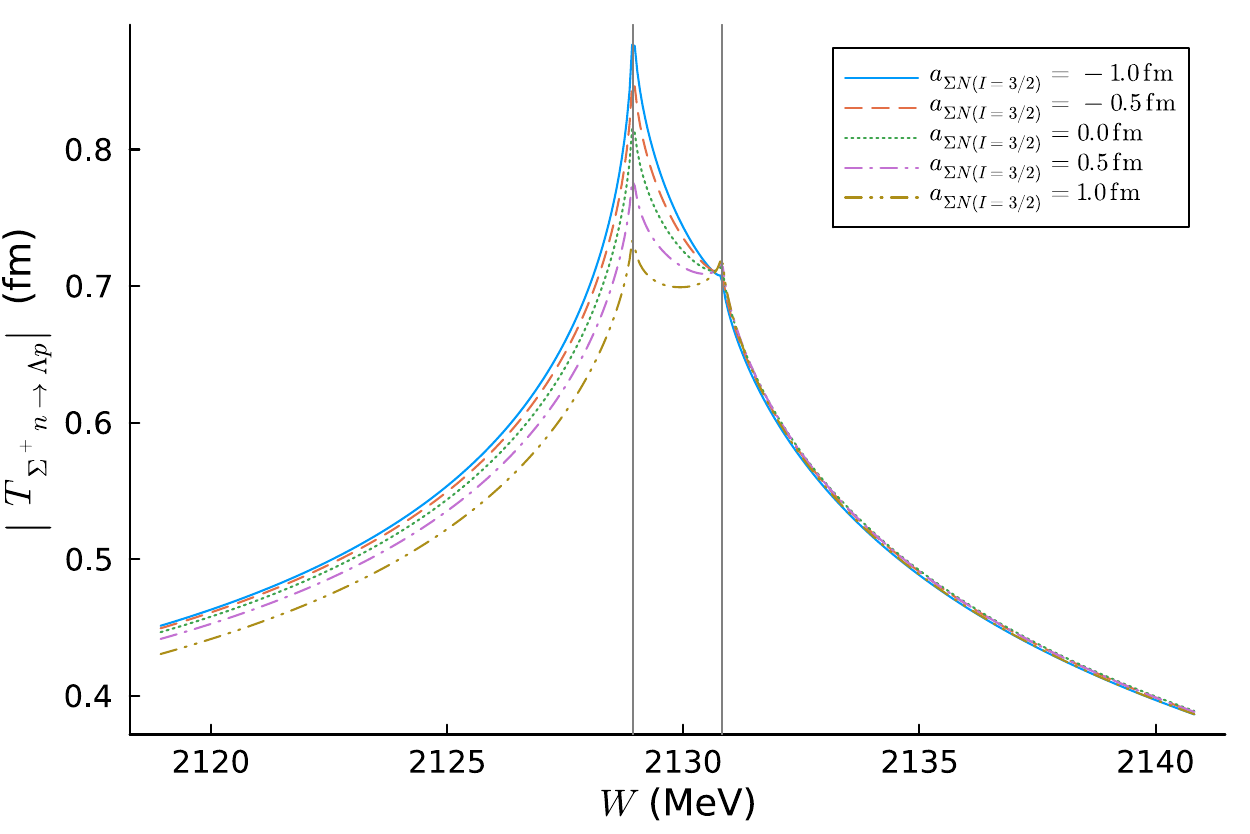}
\subcaption{Dependence on the $I=3/2$ scattering length, with $\mathrm{Re}(a_{\Sigma N(I=1/2)}) = -2.0$~fm and $\mathrm{Im}(a_{\Sigma N(I=1/2)}) = 2.0$~fm fixed.}
\end{subfigure}
\caption{The absolute value of the conversion amplitude $|T_{\Sigma^+ n \to \Lambda p}|$ as a function of the invariant mass $W$. Panel (a) shows the envelope obtained by varying the scattering lengths simultaneously within the assumed ranges, while panels (b)–(d) illustrate the individual dependence on each parameter. The vertical lines indicate the $\Sigma^+ n$ and $\Sigma^0 p$ mass thresholds.}
\label{fig:param}
\end{figure*}

\section{hyperon--nucleon amplitude}\label{sec2}
We describe the spin-triplet hyperon--nucleon ($YN$) scattering amplitude within the $K$-matrix formalism. Due to kinematic spin selection for events where the pion is emitted in the forward direction~\cite{yasunaga_2025}, only the spin-triplet final-state interaction contributes to the $\Lambda p$ invariant mass spectrum. The three-channel $YN$ system consists of the particle states $\Lambda p$, $\Sigma^0p$, and $\Sigma^+n$, and the scattering amplitude is given by
\begin{equation}
    T_{YN}(W) = \left(K^{-1}-iP(W)\right)^{-1}
\end{equation}
where $W$ denotes the invariant mass, and $P$ is the diagonal matrix of the channel momenta. The imaginary part $iP$ ensures the unitarity of the scattering amplitude. The constant $K$-matrix in the particle basis is constructed from the isospin basis as
\begin{equation}
    K = CK_\text{iso}C^T,
\end{equation}
where $C$ is the transformation matrix composed of the Clebsch--Gordan coefficients, and $K_\text{iso}$ is a symmetric constant matrix defined in the isospin basis:
\begin{align}
    C&=\begin{pmatrix}
        1&0&0\\0&\sqrt{\frac{2}{3}}&\sqrt{\frac{1}{3}}\\0&-\sqrt{\frac{1}{3}}&\sqrt{\frac{2}{3}}
      \end{pmatrix},\\
      K_\text{iso}&=\begin{pmatrix}K_{\Lambda\Lambda}&K_{\Lambda\Sigma}&0\\K_{\Lambda\Sigma}&K_{\Sigma\Sigma(I=1/2)}&0\\0&0&K_{\Sigma\Sigma(I=3/2)}\end{pmatrix}.
\end{align}
We fix four constants $K_{\Lambda\Lambda}$, $K_{\Lambda\Sigma}$, $K_{\Sigma\Sigma(I=1/2)}$, and $K_{\Sigma\Sigma(I=3/2)}$ using the scattering lengths in the isospin basis as 
\begin{align}
    \left(K_\text{iso}^{-1}-iP_\text{iso}(M_\Lambda+M_N)\right)^{-1}_{11}&=-a_{\Lambda p},\\
    \left(K_\text{iso}^{-1}-iP_\text{iso}(M_\Sigma+M_N)\right)^{-1}_{22}&=-a_{\Sigma N(I=1/2)},\\
    \left(K_\text{iso}^{-1}-iP_\text{iso}(M_\Sigma+M_N)\right)^{-1}_{33}&=-a_{\Sigma N(I=3/2)},
\end{align}
where $P_\text{iso}$ is the diagonal matrix of the channel momenta in the isospin basis. For the $\Lambda p$ system, we use an empirical value $a_{\Lambda p}=-1.56^{+0.19}_{-0.22}$~\cite{budzanowski_2010} throughout this work.

Our focus is on the $\Sigma N \to \Lambda p$ conversion amplitudes, $T_{\Sigma^+ n \to \Lambda p}$ and $T_{\Sigma^0 p \to \Lambda p}$. We first examine their dependence on the $K$-matrix parameters. Figure~\ref{fig:param} shows the absolute values of the amplitude $T_{\Sigma^+ n \to \Lambda p}$ for selected parameter sets. Panel~(a) shows the envelope, defined as the outermost boundary of the results obtained by scanning the scattering lengths over the assumed ranges: $\mathrm{Re}(a_{\Sigma N(I=1/2)}) \in [-4.0,-1.0]$~fm and $[1.0,4.0]$~fm, $\mathrm{Im}(a_{\Sigma N(I=1/2)}) \in [1.0,4.0]$~fm, and $a_{\Sigma N(I=3/2)} \in [-1.0,1.0]$~fm. All parameters are varied simultaneously within these ranges. The blue (red) band corresponds to the cases with positive (negative) values of $\mathrm{Re}(a_{\Sigma N(I=1/2)})$. Since the magnitude of $\mathrm{Re}(a_{\Sigma N(I=1/2)})$ is restricted to be larger than 1.0~fm, the two bands represent solutions with the same magnitude range but opposite signs. This figure clearly demonstrates a strong sensitivity of the amplitude to the sign of the real part of the $I=1/2$ scattering length: negative values suppress the enhancement near the $\Sigma N$ threshold, while positive values lead to a pronounced peak structure. Panels~(b)–(d) illustrate the dependence on each parameter individually. In panel~(b), $\mathrm{Re}(a_{\Sigma N(I=1/2)})$ is varied while fixing $\mathrm{Im}(a_{\Sigma N(I=1/2)}) = 2.0$~fm, demonstrating that the peak structure near the $\Sigma N$ threshold is significantly enhanced for positive values. Panel~(c) shows the dependence on the imaginary part of $a_{\Sigma N(I=1/2)}$ with $\mathrm{Re}(a_{\Sigma N(I=1/2)}) = -2.0$~fm fixed, where increasing $\mathrm{Im}(a_{\Sigma N(I=1/2)})$ leads to a broadening and enhancement of the spectrum. Finally, panel~(d) displays the dependence on $a_{\Sigma N(I=3/2)}$, indicating that its effect is relatively minor compared with the $I=1/2$ component, mainly producing small variations around the threshold.

\section{$\Lambda p$ invariant mass spectrum of the $K^-d\to\pi^-\Lambda p$ reaction}
\begin{figure}[ht]
    \centering
    \includegraphics[width=1.0\linewidth]{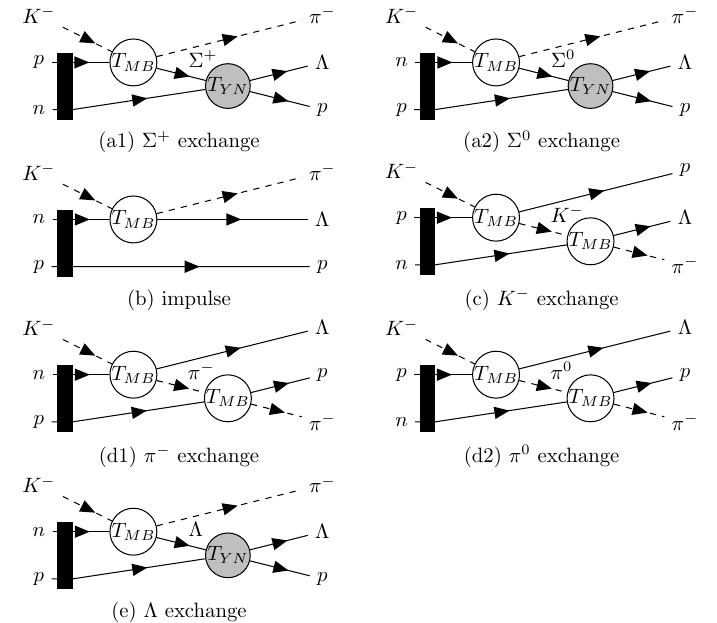}
    \caption{Feynman diagrams for the $K^-d\to\pi^-\Lambda p$ reaction.}
    \label{fig:diagram}
\end{figure}
The $\Lambda p$ invariant mass spectrum of the $K^-d\to\pi^-\Lambda p$ reaction is calculated as
\begin{equation}
\frac{d\sigma}{dM_{\Lambda p}} = \frac{M_d M_\Lambda M_p}{(2\pi)^5 4 k_{cm} E_{cm}^2}\int d\Omega d\Omega^* |\mathcal{T}|^2 p_\pi p_\Lambda^*, \label{equ:cross_section}
\end{equation}
where $k_\text{cm}$ and $E_\text{cm}$ are the center-of-mass momentum and energy of the $K^-d$ system, respectively. Here, $\Omega$ and $p_\pi$ denote the solid angle and momentum of the pion in the center-of-mass frame, while $\Omega^*$ and $p_\Lambda^*$ represent the solid angle and momentum of $\Lambda$ in the $\Lambda p$ rest frame. The quantity $|\mathcal{T}|^2$ is the spin averaged squared scattering amplitude. Details of the formulation are given in Ref.~\cite{yasunaga_2025}, and a more detailed description will be presented in a forthcoming paper~\cite{yasunaga_inprep}.

Figure~\ref{fig:diagram} shows the factorized diagrams for the reaction up to two-step processes. The $\Sigma$-exchange diagrams (a1) and (a2), which include the $\Sigma N \to \Lambda p$ conversion, are the main focus of this study. The diagrams (b), (c), (d1), (d2), and (e) represent background processes. The impulse diagram (b) provides the single-step contribution, which is dominant in the absence of kinematical selection. This contribution can be suppressed by selecting events with higher nucleon momentum in the final state. The meson-exchange diagrams (c), (d1), and (d2), as well as the $\Lambda$-exchange diagram (e), can be suppressed by selecting forward pion angles.

Here, we focus only on the contribution from the $\Sigma$-exchange diagrams.
The amplitude is constructed as the product of the elementary
$K^-N \to \pi^-Y$ amplitude and the final-state
$\Sigma N\to\Lambda p$ scattering amplitude, folded with the deuteron wave
function and the intermediate-hyperon propagator. For the $K^-N \to \pi^-Y$ amplitude, we employ the ANL-Osaka DCC model~\cite{kamano_2014},
including partial waves up to the $f$-wave ($l=3$). The amplitude is rescaled to match the mass dimension used in the present calculation and is Lorentz-transformed to the deuteron rest frame. The $\Sigma N \to \Lambda p$ amplitude is constructed as described in Sec.~\ref{sec2}, where only the $s$-wave ($l=0$) component is taken into account. For the deuteron wave function, we use the CD-Bonn potential~\cite{machleidt_2001} in momentum space, including both the $s$- and $d$-wave ($l=2$) components. In the present calculation, the input parameters for the $K$-matrix of $YN$ amplitude are taken from the scattering lengths predicted by several theoretical models~\cite{rijken_1999,haidenbauer_2005,haidenbauer_2023}. As summarized in Table~\ref{tab:param}, we employ three parameter sets for the spin-triplet $\Sigma N$ scattering lengths in the $I=1/2$ and $I=3/2$ channels. 

\begin{table}[htbp]
    \centering
    \begin{tabular}{c|cc}
        &$a_{\Sigma N(I=1/2)}$ [fm]&$a_{\Sigma N(I=3/2)}$ [fm]\\\hline
        Param.~1~\cite{rijken_1999}&$1.68-2.35i$ &$-0.25$\\
        Param.~2~\cite{haidenbauer_2005}&$-3.83-3.01i$ &$0.29$\\
        Param.~3~\cite{haidenbauer_2023}&$2.53-2.64i$ &$0.41$
    \end{tabular}
    \caption{Some parameters}
    \label{tab:param}
\end{table}

\begin{figure}
    \centering
    \includegraphics[width=1.0\linewidth]{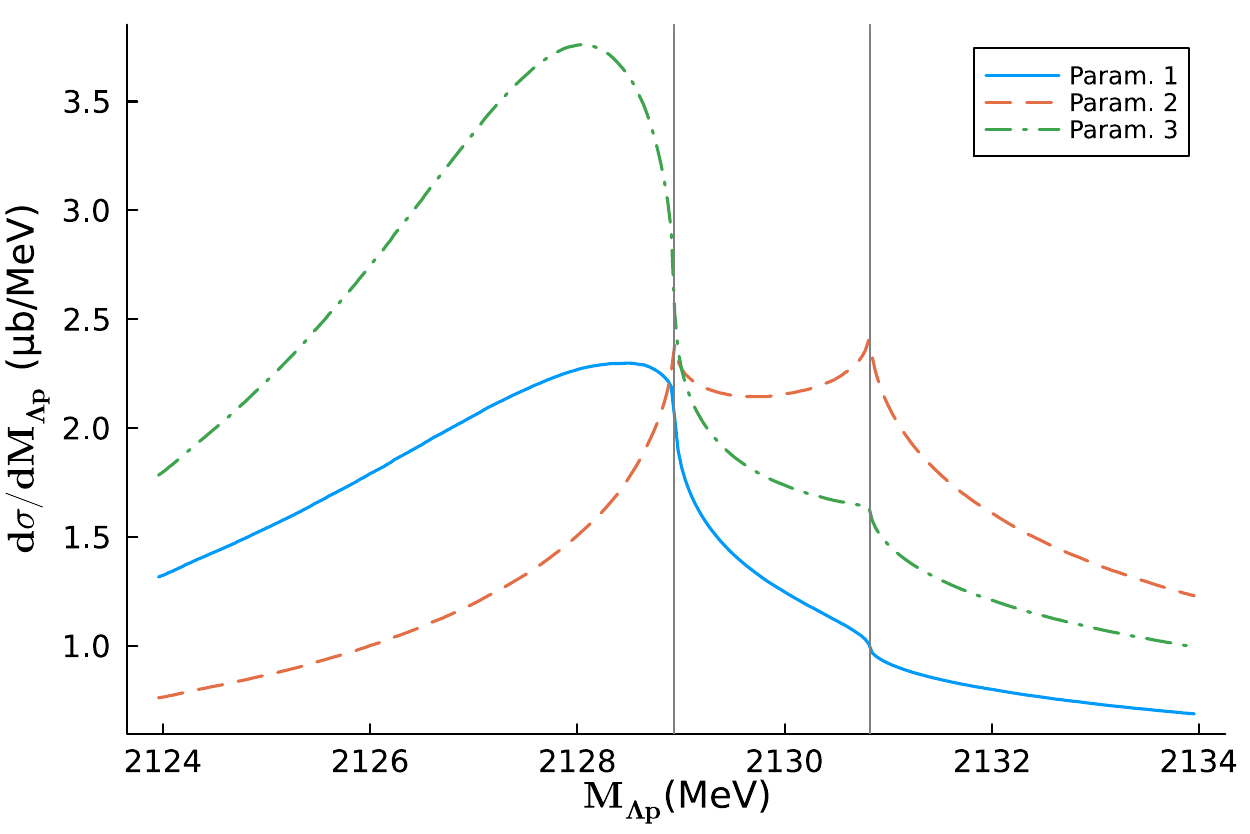}
    \caption{$\Lambda p$ invariant mass spectra of the $K^-d\to\pi^-\Lambda p$ reaction calculated with three different parameter sets for the $\Lambda N$--$\Sigma N$ coupled-channel amplitude. The vertical lines indicate the $\Sigma^+ n$ and $\Sigma^0 p$ thresholds.}
    \label{fig:spectrum}
\end{figure}

Figure~\ref{fig:spectrum} shows the $\Lambda p$ invariant mass spectra
calculated with three different parameter sets for the $\Lambda N$--$\Sigma N$
coupled-channel amplitude. The initial kaon momentum is $1.0$~GeV/$c$, and the contributions from background diagrams are included. To reduce the background contributions, we impose kinematical cuts of $\cos\theta_\pi>0.9$ for the pion angle and $p_n>200$~MeV/$c$ for the neutron momentum. In all cases, characteristic structures appear around the $\Sigma N$ thresholds. The shape of the spectrum depends on the choice of the interaction parameters. In particular, Param.~1 and Param.~3 show an enhancement below the $\Sigma^+ n$ threshold, followed by a decrease above the threshold. On the other hand, Param.~2 exhibits a sharp enhancement near the $\Sigma^+ n$ and $\Sigma^0 p$ threshold regions. These differences originate from the sign of the real part of the $I=1/2$ scattering length, as confirmed in Fig.~1.

\section{Summary}
In this work, we have investigated the $\Lambda N$--$\Sigma N$ coupled-channel interaction through the $\Lambda p$ invariant mass spectrum of the $K^-d\to\pi^-\Lambda p$ reaction. The spin-triplet $\Sigma N\to\Lambda p$ conversion amplitude has been constructed within the $K$-matrix formalism using scattering lengths in the isospin basis. We have examined the parameter dependence of the conversion amplitude and found that the structure near the $\Sigma N$ threshold is particularly sensitive to the sign of the real part of the $I=1/2$ scattering length. We have also calculated the $\Lambda p$ invariant mass spectrum from the $\Sigma$-exchange diagrams.
Characteristic threshold structures have been found around the $\Sigma N$ thresholds, and their shapes have been shown to depend strongly on the choice of the interaction parameters. These results suggest that the $\Lambda p$ invariant mass spectrum in the $K^-d\to\pi^-\Lambda p$ reaction can serve as a useful observable for investigating the $\Lambda N$--$\Sigma N$ coupled-channel interaction.

\section*{Acknowledgments}
S. Y. was partly supported by JST SPRING Grant No. JPMJSP2180. D. J. was partly supported by Grants-in-Aid for Scientific Research No.~JP22H04917, No.~JP23K03427, and No.~JP25K07315. We also used the Yukawa-21 supercomputer at the Yukawa Institute for Theoretical Physics, Kyoto University.

\bibliography{reference}

\end{document}